\documentstyle[epsfig,aps,floats,preprint,tighten]{revtex}

\def\beq{\begin{equation}}
\def\eeq{\end{equation}}

\def\etal{{\it et al.}}
\def\epsbol{\mbox{\boldmath$\epsilon$}}

\def\sigbol{\mbox{\boldmath$\sigma$}}

\def\N{{\scriptscriptstyle N}}
\def\A{{\scriptscriptstyle A}}

\def\eabcd{\epsilon_{\alpha\beta\gamma\delta}}

\begin{document}
\draft
\preprint{MC/TH 99/12}

\title{Spin polarisability of the nucleon at NLO in the chiral expansion}
\author{K. B. Vijaya Kumar\footnote{Permanent address: Department of Physics,
Mangalore University 574 199, India}, 
Judith A. McGovern\footnote{Electronic address: judith.mcgovern@man.ac.uk} 
and Michael C. Birse\footnote{Electronic address: mike.birse@man.ac.uk} }
\vskip 20pt
\address{Theoretical Physics Group, Department of Physics and Astronomy\\
University of Manchester, Manchester, M13 9PL, U.K.}
\nopagebreak
\maketitle
\begin{abstract}
We present a calculation of the fourth-order (NLO) contribution to 
spin-dependent forward Compton scattering in heavy-baryon chiral perturbation 
theory. No low-energy constants, except for the anomalous magnetic moments of 
the nucleon, enter at this order. The fourth-order piece of the spin 
polarisability of the proton turns out to be almost twice the size of the 
leading piece, with the opposite sign. This leads to the conclusion that no 
prediction can currently be made for this quantity.

\end{abstract}
Compton scattering from the nucleon has recently been the subject of much work,
both experimental and theoretical. For the case of unpolarised
protons the experimental amplitude is well determined, and in good agreement
with the results of heavy-baryon chiral perturbation theory (HBCPT).
However the situation with regard to scattering  from polarised targets is less
satisfactory, not least because until very recently no direct measurements of
polarised Compton scattering had been attempted.

The usual notation for spin-dependent pieces of the forward scattering 
amplitude for real photons of energy $\omega$ and momentum $q$ is
\beq
\epsilon_1^\mu\Theta_{\mu\nu} \epsilon_2^\nu=i e^2 \omega W^{(1)}(\omega)
\sigbol\cdot(\epsbol_1\times\epsbol_2)+\ldots
\label{amp}
\eeq
From a theoretical perspective there is particular interest in the low-energy 
limit of the amplitude:
 $e^2 W^{(1)}(\omega)=4\pi(f_2(0)+\omega^2\gamma)+\ldots$, where
$\gamma$ is the forward spin-polarisability.
The low-energy theorem (LET) of Low, Gell-Mann and Goldberger \cite{LGG}
states that $f_2(0)=-\alpha_{em}\kappa^2/2 M_\N^2$.

In terms of measurable quantities, the low-energy constants 
$f_2(0)$ and $\gamma$ can be obtained from measurements at energies above 
the threshold for pion production, $\omega_0$, via
dispersion relations;
that for the polarisability gives
\beq
\gamma={1\over 4\pi^2}
\int^\infty_{\omega_0}{\sigma_{-}(\omega)-\sigma_{+}(\omega)\over
\omega^3}d\omega, 
\eeq
where $\sigma_{\pm}$ are the parallel and antiparallel cross-sections
for photon absorption; 
the related sum rule for $f_2(0)$, due to Gerasimov Drell and Hearn, 
\cite{GDH} has the same form except that $1/\omega$ replaces $1/\omega^3$.

Before direct data existed, the relevant cross-sections were estimated  from
multipole analyses of pion electroproduction  experiments \cite{karl,sand}. 
These showed significant discrepancies between the LET and the GDH sum rule for
the  difference of $f_2(0)$ for the proton and neutron, though the sum was in
good agreement.  Indeed even the sign of the difference was different. 
More recently, measurements have been made with MAMI at Mainz,
for photon energies between 200 and 800~MeV; the range will be extended
downward to 140~MeV, and a future experiment at Bonn will extend it upwards to
3~GeV \cite{thomas}.   
The preliminary data from MAMI \cite{thomas} suggest a continuing discrepancy
between the LET and the sum rule for the proton, though a smaller one than
given by the multipole analysis. The most recent analysis using
electroproduction data, which pays particular attention to the threshold
region, also reduces the discrepancy somewhat \cite{krein1}.

The MAMI data does not currently go low enough in energy to give a reliable 
result for the spin polarisability, $\gamma$.  However electroproduction data
have also been  used to extract this quantity; Sandorfi \etal  \cite{sand} 
find 
$\gamma_p=-1.3 \times 10^{-4}$~fm$^4$ and $\gamma_n=-0.4 \times
10^{-4}$~fm$^4$, while the more recent analysis of Drechsel \etal \cite{krein2}
gives a rather smaller value of $\gamma_p=-0.6 \times 10^{-4}$~fm$^4$. 
(We shall use units of $10^{-4}$~fm$^4$ for polarisabilities from now on.)
                                                                 
The spin polarisability has also been calculated in the framework of 
HBCPT: at lowest (third) order in the chiral expansion this gives
$\gamma=\alpha_{em}g_\A^2/(24 \pi^2 f_\pi^2 m_\pi^2)=4.54$
for both proton and neutron, where the entire contribution comes from 
$\pi N$ loops. The effect of the $\Delta$ enters in counter-terms at fifth 
order in standard HBCPT, and has been estimated to be so large as to change the 
sign \cite{ber92}.
The calculation has also been done in an extension of HBCPT with an explicit 
$\Delta$ by Hemmert \etal \cite{hemm}.  They find that the principal effect is
from the $\Delta$ pole, which contributes $-2.4$, with the effect of
$\pi\Delta$ loops being small, $-0.2$.  Clearly the next most important
contribution is likely to be the fourth-order $\pi N$ piece, and this is the
result which is presented  here.  The effects of the $\Delta$ at NLO involve
unknown parameters; we might hope that the loop pieces at least will be small.

\begin{figure}
  \begin{center} \mbox{\epsfig{file=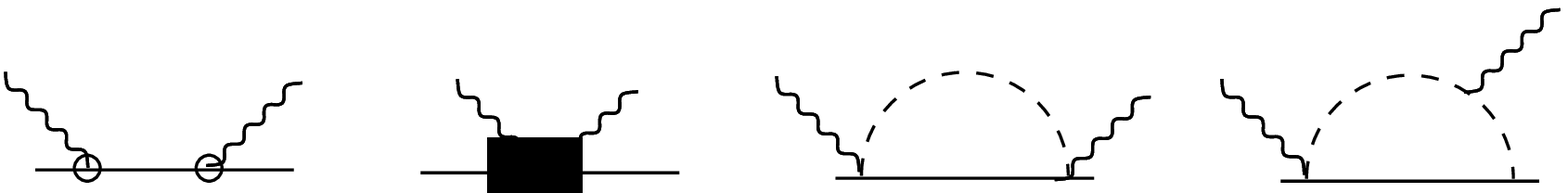,width=16truecm,angle=0}}
  \end{center}
{\bf Fig.~1:}
Diagrams which contribute to spin-dependent Compton scattering in the 
$\epsilon\cdot v=0$ gauge at LO. The open circles are vertices
from ${\cal L}^{(2)}$ and the solid dot is a vertex from ${\cal L}^{(3)}$.
\end{figure}

To calculate the spin-dependent forward scattering amplitude,
we work in the gauge $A_0=0$, or in the language of HBCPT, $v\cdot\epsilon=0$,
where $v^\mu$ is the unit vector which defines the nucleon rest frame. 
The amplitude of Eq.~(\ref{amp}) becomes
\beq
\epsilon_1^\mu\Theta_{\mu\nu} \epsilon_2^\nu=
2i e^2 v\cdot q W^{(1)}(v\cdot q) \eabcd v^\alpha \epsilon_1^\beta
\epsilon_2^\gamma S^\delta+\dots
\eeq
where $S^\mu$ is the spin operator, which obeys $v\cdot S=0$.
The LET states that $W^{(1)}(0)=-e^2\kappa^2/2M_\N^2$,
where $\kappa$ is the anomalous magnetic moment of the nucleon.
At leading (third) order this is satisfied, with $\kappa$ replaced by its
bare value, by the combination of the Born terms and the seagull diagram,
which has a fixed coefficient in the third-order Lagrangian \cite{ber92}.
The loop diagrams of figure 1 have contributions of order $\omega$ which 
cancel and so do not affect the LET. At order $\omega^3$ they 
give the result quoted above for the polarisability.  Note that in
this gauge there is no lowest-order coupling of a photon to a nucleon; the 
coupling comes in only at second order. The Feynman vertex consists of two
pieces, one proportional to the charge current and one to the magnetic moment:
\beq
{ie\over 2M}\Bigl(Q\epsilon\cdot(p_1+p_2)+2(Q+\kappa)[S\cdot\epsilon,S\cdot
q]\Bigr).
\eeq
This and all other vertices are taken from the review of Bernard 
\etal \cite{mei95}. The two pieces contribute in different diagrams; the first 
is represented by a solid dot and the second by an open circle in 
figures 1 and 2.

\begin{figure}
  \begin{center} \mbox{\epsfig{file=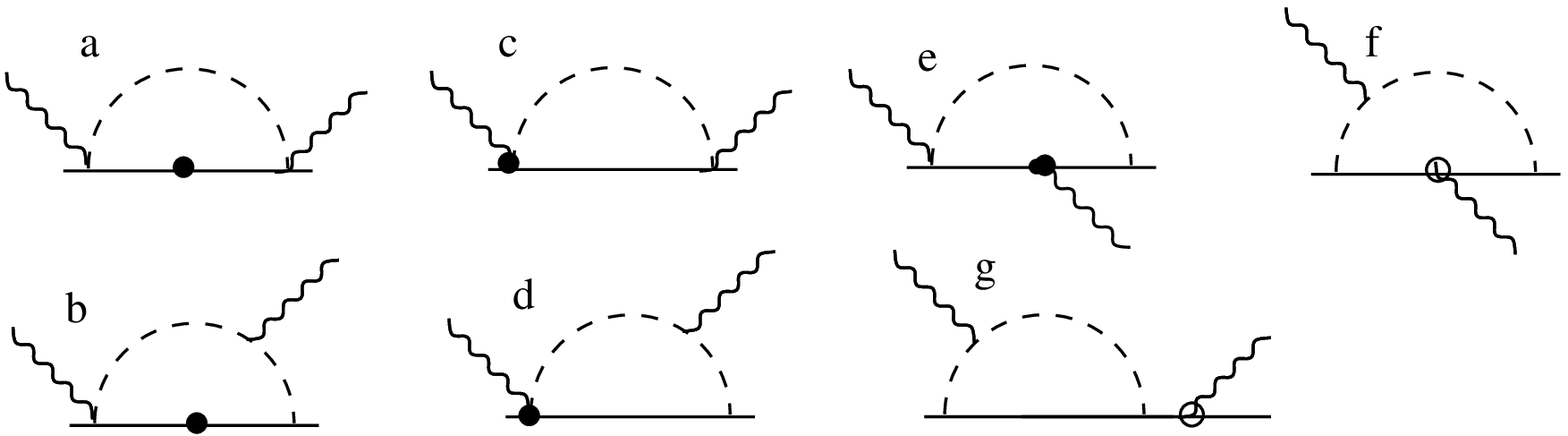,width=16truecm,angle=0}}
  \end{center}
{\bf Fig.~2:}
Diagrams which contribute to spin-dependent forward Compton
scattering in the $\epsilon\cdot v=0$ gauge at NLO. The dots and circles are
vertices from ${\cal L}^{(2)}$, the solid and open dots at the 
photon-nucleon vertices representing the couplings proportional to the charge 
current and magnetic moment respectively.
\end{figure}

At NLO, the diagrams which contribute are given in figure 2.  There can be
no seagulls at this order; since $W^{(1)}(\omega)$ is of first chiral order
and is even in $\omega$ (two powers of $e$ and one of $\omega$ having been 
pulled out of the amplitude in its definition), it will have an expansion of
the form $a m_\pi +b\omega^2/m_\pi+\ldots$. 
These non-analytic powers of $m_\pi^2$ cannot be present in the basic
couplings in the Lagrangian, but can only be generated from loops.  It follows 
that there are no undetermined low-energy constants in the final amplitude.

The insertion on the nucleon propagator of figures 2a and b needs 
some explanation.  Denoting the external nucleon residual momentum by $p$,
the energy $v\cdot p$ starts at second chiral order with the mass shift and
kinetic energy. In contrast the space components of $p$ and all components
of the loop momentum $l$ and the photon  momentum are first order. The
propagator with an insertion  consists of both the second term in the expansion
of the lowest-order  propagator, $i/(v\cdot l+v\cdot p)$, in powers of $v\cdot
p/v\cdot l$, and also the insertions from ${\cal L}^{(2)}$. The second-order
mass shift and external kinetic energy cancel between the two to leave, in the
rest frame, just the second order propagator given in appendix A of 
ref.~\cite{mei95}.

The contributions of the various diagrams from figure 2 are given in table 1.
The final result has  the following form:
\beq
W^{(1)}_4(w)={g_\A^2m_\pi\over 48 \pi^2 f_\pi^2 M_\N x^2}
\left(A+B x \,\hbox{arccos}^2(-x)
+{(C+Dx^2+Ex^4)\hbox{arccos}(-x)\over \sqrt{1-x^2}}\right)+(x\to-x)
\label{fullform}
\eeq
where $x=\omega/m_\pi$ and the constants $A-E$ are given below:
\begin{eqnarray}
A&=&-\pi\bigl(1-\kappa_v +(2+\kappa_s)\tau_3\bigr)\\
B&=&3\kappa_s\tau_3/2\\
C&=&2\bigl(1-\kappa_v+(2+\kappa_s)\tau_3\bigr)\\
D&=&-1+4\kappa_v-(2+\kappa_s)\tau_3\bigr)\\
E&=&-\bigl(4+2\kappa_v+(2+\kappa_s)\tau_3\bigr)
\end{eqnarray}
It is worth pointing out that the Born-like contribution 2g, like all
the two-particle irreducible loop diagrams, is perfectly finite as $\omega\to
0$, as the momentum-dependence of the magnetic coupling cancels the $1/\omega$
of the propagator.  Thus the expression in Eq.~(\ref{fullform}) is also finite,
despite superficial appearances.

The total contribution at order $\omega^0$ is 
\beq W^{(1)}_4(0)={g_\A^2m_\pi\over 16\pi M_\N f_\pi^2}
(\kappa_v+\kappa_s\tau_3),
\eeq
which, since the one-loop contribution to $\kappa$ is 
$\delta \kappa_v= -g_\A^2 m_\pi M_\N/4 \pi f_\pi^2$, 
can be seen to be exactly the correction to the leading contribution required 
to satisfy the LET.

The polarisability to NLO is
\beq
\gamma={\alpha_{em}g_\A^2 \over 24\pi^2 f_\pi^2 m_\pi^2}\left[1 
-{\pi m_\pi \over 8 M_\N}\bigr(15+3\kappa_v+(6+\kappa_s)\tau_3\bigr)\right].
\eeq
Although this has a factor of $m_\pi/M_\N$ compared with the leading piece,
the numerical coefficient is large.  Using the physical values of the masses 
and anomalous magnetic moments gives $\gamma=4.5-(6.9+1.6\,\tau_3)$.  
The NLO contributions are
disappointingly large, and call the convergence of the expansion into question.

\begin{figure}
\def\w2{\omega^2}
\def\m2{m^2}
\begin{center}
\begin{tabular}{|c|c|c|c|c|} \hline
Diagram &isospin & $O(\omega)$ & $O(\omega^3)$  & full function: odd piece
only \\ \hline\hline
a  & 1 &$3$ &$-5/2$ & $-(2\w2-\m2)\partial J_0(\omega)/\partial \omega-
                                   2\omega J_0(\omega)$ \\ \hline
b & 1 &$-5$ & $  7/ 4$ &$-4\Bigl(\partial J_2(\omega)/\partial\omega+ 
                               m^2 (J'_2(\omega)-J'_2(0))/\omega\Bigr) $\\ \hline
c & $1-\tau_3$ &1 &$-1/ 4$ & $2(J_2(\omega)-J_2(0))/\omega$ \\ \hline
d & $\tau_3$ &$- 2 $&$1 /3$ & $4\omega\int^1_0J'_2(x \omega) dx$\\ \hline
e  & $\tau_3$ &2 & $-1$ & $-2\omega J_0(\omega)$\\ \hline
f & $1+\kappa_v-(1+\kappa_s)\tau_3 $ & 0 & $ -1/ 12$ 
                        & $2\omega\int^1_0(1-2x)J'_2(x \omega) dx$\\ \hline
g & $1+\kappa_v+(1+\kappa_s)\tau_3 $ & 1 &$ -1/ 6$ &
                          $-2\omega\int^1_0J'_2(x \omega) dx$\\ \hline
\end{tabular}
\end{center}                       
{\bf Table 1:}
Contributions to $2\omega W^{(1)}(\omega)$ from the diagrams of figure 2.
All three columns are multiplied by a common factor of $g_\A^2/(M_\N f^2)$; the
coefficients  of $\omega$ and $\omega^3$ have an extra factor of $m_\pi/8\pi$.
The $J_i$ are defined in ref.~\cite{mei95}; a prime denotes differentiation 
with respect to $m^2$.  
\end{figure}

When this work was nearly complete, a preprint on spin-dependent Compton 
scattering at NLO was sent to the archives by Ji \etal
\cite{osborne}. They chose not to work in the $v\cdot\epsilon=0$ gauge, since 
it is not convenient for the extension to virtual photons which forms their 
main interest. The price of this is many more diagrams to be evaluated,
and two extra Lorentz structures appear multiplying $W^{(1)}$ (namely 
$S\cdot(\epsilon_1\times q) v\cdot \epsilon_2 -(\epsilon_1 \leftrightarrow
\epsilon_2)$). Some of these diagrams have poles at $\omega = 0$.  
However Ji and coworkers have not included diagrams which contribute only pole
terms \cite{ji}. As a result their Eq.~(10) is incomplete, but after
subtracting the pole it agrees with our expression  Eq.~(\ref{fullform}); in
particular their expression for $\gamma$ agrees with ours. 

JMcG and MCB acknowledge the support of the UK EPSRC. VK holds a Commonwealth
Fellowship.

\nopagebreak

\end{document}